\documentclass[twocolumn]{revtex4-1}
\usepackage{graphics,epsfig,amsfonts,amssymb,amsmath,color}
\usepackage[normalem]{ulem}

\begin{document}

\title{Interface effects on acceptor qubits in silicon and germanium}
\author{J.C. Abadillo-Uriel}
\email{jcgau@icmm.csic.es}
\affiliation{Instituto de Ciencia de Materiales de Madrid, ICMM-CSIC, Cantoblanco, E-28049 Madrid (Spain)}
\author{M.J. Calder\'on}
\affiliation{Instituto de Ciencia de Materiales de Madrid, ICMM-CSIC, Cantoblanco, E-28049 Madrid (Spain)}
\date{\today}
\begin{abstract}
Dopant-based quantum computing implementations often require the dopants to be situated close to an interface to facilitate qubit manipulation with local gates. Interfaces not only modify the energies of the bound states but also affect their symmetry. Making use of the successful effective mass theory we study the energy spectra of acceptors in Si or Ge taking into account the quantum confinement, the dielectric mismatch and the central cell effects. The presence of an interface puts constraints to the allowed symmetries and lead to the splitting of the ground state in two Kramers doublets [J. Mol et al, App. Phys. Lett. {\bf 106}, 203110 (2015)]. Inversion symmetry breaking also implies parity mixing which affects the allowed optical transitions.  Consequences for acceptor qubits are discussed. 
\end{abstract}
\maketitle

\section{Introduction}

Dopants, carrier providers for traditional transistors, are acquiring a more important role as semiconductor nanostructures shrink their size. This change of paradigm was envisioned a few years ago with the first proposal for the implementation of a scalable dopant based quantum computer~\cite{KaneNature1998}. Silicon is an excellent platform for this technology due to its resilience against decoherence and the already existing high level control of Si nanoelectronics~\cite{ZwanenburgRevModPhys2013}. Proposals relied originally on the quantum control of single electrons bound to donors, being the electron spin~\cite{VrijenPRA2000}, electron charge~\cite{BarrettPRB2003} or the nuclear spin~\cite{KaneNature1998} the degrees of freedom used as qubits. Recent years have witnessed the practical demonstration of some of the ingredients involved in these proposals including single-shot spin readout of bound electrons~\cite{MorelloNat2010}, exchange coupling between electrons bound to neighbouring donors~\cite{ZalbaNanoLett2014,DehollainPRL2014} and the practical implementation of the electrical control of spin qubits by an A-gate~\cite{LauchtScienceAdv2015} as proposed in Ref.~\cite{KaneNature1998}. 

Proposals of acceptor-based qubits may make use of the long range strong dipolar inter-qubit coupling~\cite{GoldingArXiv2003}, or exploit the spin-orbit interaction (which is stronger than for electrons) to couple spin to phonons~\cite{RuskovPRB2013}, or to oscillating electric fields~\cite{SalfiArXiv2015}.  The relative importance of the different sources of decoherence is different in electrons and holes: spin-orbit interactions~\cite{HuangPRB2014,BermeisterAPL2014} would be more important for holes while hyperfine interaction (which can cause spin decoherence due to coupling to nuclear spins~\cite{WitzelPRB2006}) is smaller for holes than for electrons~\cite{ChekhovichNatPhys2013}. The effective suppression of the latter by Si isotopic purification, which gets rid of nuclear spins in Si, leads to very long electron coherence times~\cite{TyryshkinNatMat2012}. 

\begin{figure}
\includegraphics[clip,width=0.23\textwidth]{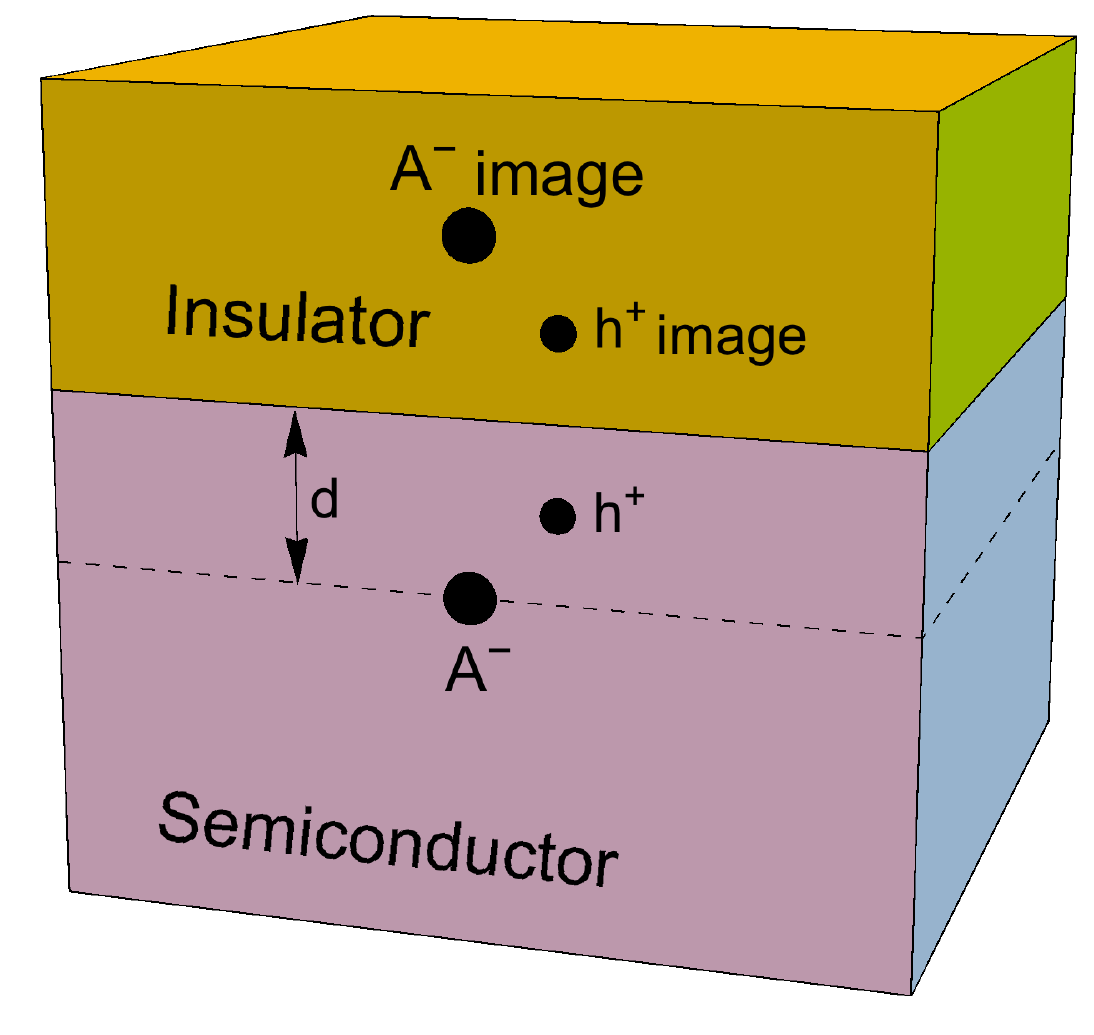}
\includegraphics[clip,width=0.23\textwidth]{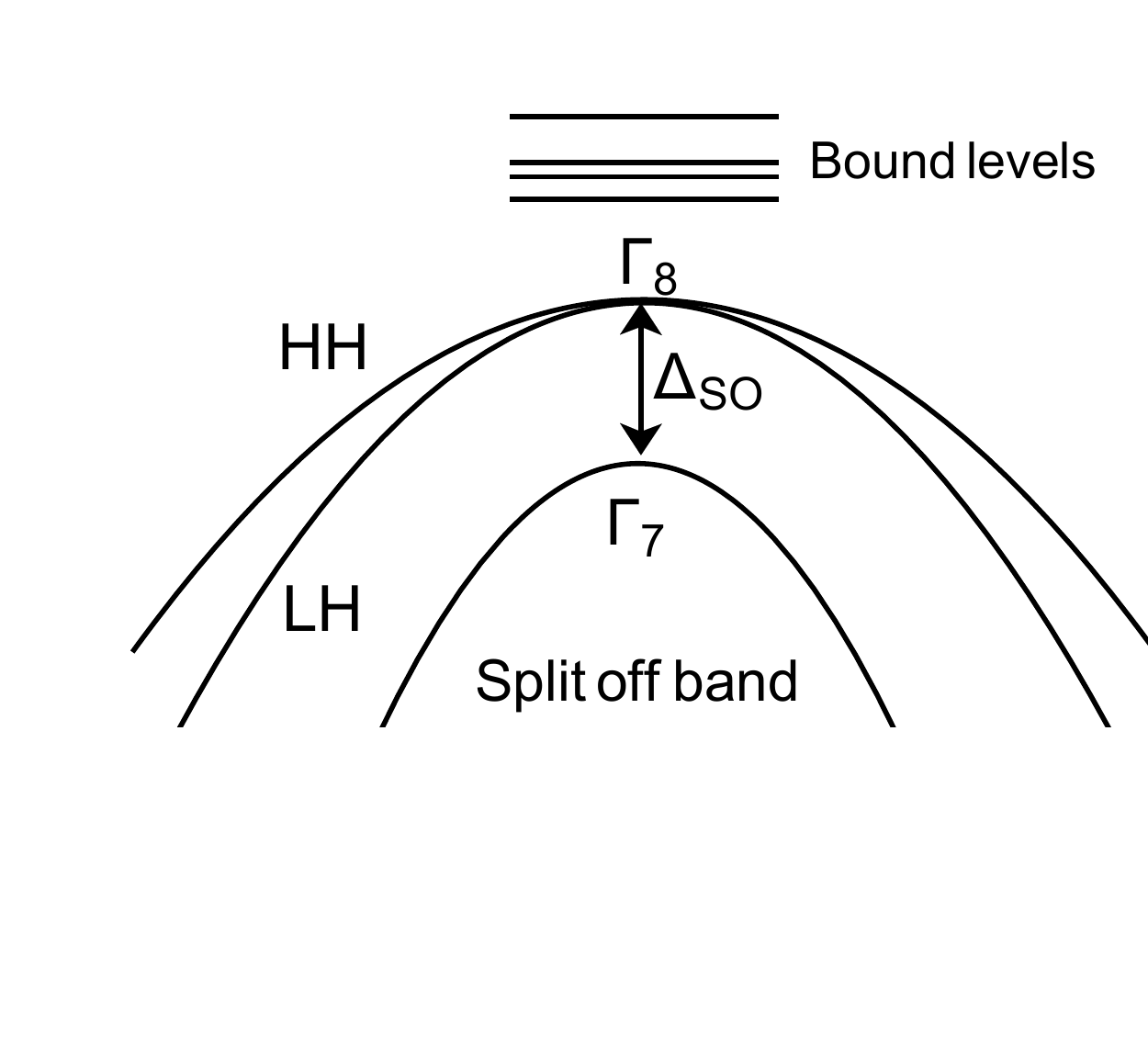}
\caption{\label{fig:problem} Left: Schematic view of the problem. The acceptor A is at a distance $d$ from the $(001)$ interface between the semiconductor (Si or Ge) and an insulating barrier. Image charges appear due to the dielectric mismatch between the semiconductor and the barrier. Right: Sketch of the valence subbands. HH indicates the heavy hole subband while LH indicates the light hole subband. The energies of the bound states are positive and defined with respect to the top of the valence band.}
\end{figure}

In practical dopant-based quantum computer proposals, dopants are often introduced in nanostructures and close to surfaces or interfaces with materials different from the host. In this case, the energies of bound carriers can be shifted by quantum confinement and dielectric mismatch~\cite{delerue-lannoo,CalderonPRB2010,MolPRB2013} potentially modifying the working parameters of the devices. Quantum confinement may alter the shape of the wave-functions through the boundary conditions, consequently affecting the binding energy. For instance, in a very thin (compared to the bound state wave-function size $a_B$) nanowire, the extra confinement enhances the binding energy deactivating the dopants as carrier providers~\cite{GarnettPRB1984,BjorkNatNano2009}. However, when the dopant is close (compared to $a_B$) to one interface/surface but not confined in other directions, the wave-function can be deformed in such a way that the density probability of the bound state decreases on the dopant, leading to a reduction of the binding energy.  The dielectric mismatch gives rise to image charges which, depending on the relative size of the dielectric functions of the nanostructure components, can lead to an enhanced or decreased binding. In the case of a semiconductor surrounded by insulators, the image charges have the same sign as the charges originating them, enhancing the binding energies. Not only the energies but also the symmetry of the bound states may be modified. In the case of acceptors, the four-fold degeneracy of the ground state may be broken by strain~\cite{BirWiley1974,BroeckxPRB1987,WangJPCM2009}, electric field~\cite{SmitPRB2004,CalvetPRL2007} and/or magnetic fields. Characterisation of dopants embedded in nanostructures can be performed via transport measurements~\cite{SellierPRL2006,PierreNatNano2010,ZwanenburgRevModPhys2013} while Scanning Tunnel Microscopy can give information of the wave-function of subsurface dopants~\cite{yakuninPRL2004,SalfiNatMat2014,MolAPL2015,SaraivaArXiv2015}.

Here we perform an analysis of the energy spectrum and the symmetry of the bound states for substitutional acceptors (group III elements) in Si or Ge close to an interface with an insulating barrier. Our study is based on effective mass theory (EMT). EMT, despite its apparent simplicity, has been proven to give very accurate descriptions not only of the binding energies but also of the donor wave functions~\cite{SaraivaArXiv2015}. EMT exploits the analogy with free atoms but includes information about the host crystal through the bands. Both quantum confinement and dielectric mismatch are included. Central cell corrections~\cite{LipariSSC1980,SaraivaJPCM2015} are also considered in order to reproduce the energy spectra of different acceptor species. A recent publication~\cite{MolAPL2015} has demonstrated that the four-fold degeneracy of the ground state is broken for B acceptors in close proximity ($\sim$ nm) to a surface. We study the energy spectra of acceptors in Si and Ge, including the observed Kramers doublet splitting of the ground state. Our approach allows a complete analysis of the symmetry breaking induced by the interface. Besides a symmetry reduction analogous to the one produced by uniaxial strain~\cite{BroeckxPRB1987}, the interface also breaks inversion symmetry, leading to parity mixing. The consequences for an acceptor based qubit are analysed.

\section{Model}
\label{sec:model}

The energy levels of a single acceptor are mainly determined by the valence band structure of the group IV host crystal, see Fig.~\ref{fig:problem}. The valence bands have their maximum at the $\Gamma$ point and the main atomic contribution belongs to $p$ states.  This implies that the total angular momentum of the bands can be $J=3/2$ or $J=1/2$. $J=3/2$ corresponds to the $\Gamma_8^{+}$ bands. At $k=0$ these bands are four-fold degenerate but split away from the $\Gamma$ point into two doubly degenerate bands: The heavy-hole bands (with $|m_J|=3/2)$ and the light-hole bands (with $|m_J|=1/2)$. The $\Gamma_7^{+}$ band corresponds to $J=1/2$ and it splits from the $\Gamma_8^{+}$ bands by the spin-orbit coupling $\Delta_{SO}$. 

The Kohn-Luttinger Hamiltonian~\cite{KohnPR1955} describes a positive hole bound to an acceptor centre taking into account the six valence subbands. The Hamiltonian  is a $6 \times  6$ matrix operator which accounts for all the possible combinations of the quantum numbers $J$ and $J_z$. 
In Si, the spin-orbit coupling is $\Delta_{SO}=44$ meV~\cite{madelungbook} which is comparable to the energy of the acceptor ground state ($\gtrsim 45$ meV), and hence the contribution of the split-off band can not be neglected. This is not the case for Ge, where the split-off energy is $290$ meV~\cite{madelungbook}, much larger than the binding energy ($\sim 10$ meV), allowing in principle the reduction of the dimensionality of the matrix operator to a $4 \times 4$ matrix. In the following we will keep the full $6 \times 6$ Hamiltonian for both Si and Ge for completeness.

In bulk, it is possible to reformulate the Kohn-Luttinger Hamiltonian separating the spherical symmetric terms, like the Coulomb impurity potential, from the terms with the cubic symmetry of the crystal~\cite{BaldereschiPRB1971, BaldereschiPRB1974, LipariSSC1978}. This allows an easier analysis of the symmetries and selection rules of the acceptor states. However, this simplification is not possible in the presence of an interface due to the reduction of the symmetry, as explained in Sec.~\ref{sec:symmetries}. Hence we keep the original $6 \times 6$ Kohn-Luttinger Hamiltonian~\cite{KohnPR1955}
\begin{widetext}
\begin{equation}
H_{\rm KL}=\begin{pmatrix}
P+Q & L & M & 0 & \frac{i}{\sqrt{2}}L & -i\sqrt{2}M \\
L^*  &  P-Q & 0 & M & -i\sqrt{2}Q & i\sqrt{\frac{3}{2}}L \\
M^* & 0 & P-Q & -L & -i\sqrt{\frac{3}{2}}L^* & -i\sqrt{2}Q \\
0  & M^* & -L^* & P+Q & -i\sqrt{2}M^* & -\frac{i}{\sqrt{2}}L^* \\
-i\sqrt{2}L^* & i\sqrt{2}Q & i\sqrt{\frac{3}{2}}L & i\sqrt{2}M & P+\Delta_{SO} & 0 \\
i\sqrt{2}M^*  & -i\sqrt{\frac{3}{2}}L^* & i\sqrt{2}Q & i\sqrt{2}L & 0 & P+\Delta_{SO}
\end{pmatrix} \,.
\label{HKL}
\end{equation}
\end{widetext}

Defining the effective Rydberg unit as $Ry^*=e^4m_0/2\hbar^2\epsilon_s^2\gamma_1$ and the effective Bohr radius as $a^*=\hbar^2\epsilon_s\gamma_1/e^2m_0$, the differential operators in Eq.~(\ref{HKL}) are 
\begin{equation}
P=-k^2+\frac{2}{r} \nonumber
\end{equation}
\begin{equation}
Q=-\frac{\gamma_2}{\gamma_1}(k_x^2+k_y^2-2k_z^2) \nonumber
\end{equation}
\begin{equation}
L=i2\sqrt{3}\frac{\gamma_3}{\gamma_1}(k_x-ik_y)k_z \nonumber 
\end{equation}
\begin{equation}
M=-\sqrt{3}\frac{\gamma_2}{\gamma_1}(k_x^2-k_y^2)+i2\sqrt{3}\frac{\gamma_3}{\gamma_1}k_x k_y  \, ,
\label{PQLM}
\end{equation}
with $m_0$ the free electron mass, $\epsilon_s$ the semiconductor static dielectric constant, and  $\gamma_1$, $\gamma_2$ and $\gamma_3$ material dependent Luttinger parameters~\cite{KohnPR1955} related to the curvature of the valence subbands. 
For Si $Ry^*(\rm Si)=24.8$ meV, $a^*(\rm Si)=2.55$ nm, $\gamma_1(\rm Si) =4.27$, $\gamma_2(\rm Si) =0.32$, $\gamma_3(\rm Si)=1.458$, while for Ge $Ry^*(\rm Ge)=4.4$ meV, $a^*(\rm Ge)=10.85$ nm,  $\gamma_1(\rm Ge) =13.35$, $\gamma_2(\rm Ge)=4.25$, $\gamma_3(\rm Ge)=5.69$~\cite{PajotSpringer2010}.

In order to account for the dependence on the acceptor species of the binding energies, the so-called central cell corrections have to be included~\cite{LipariSSC1980}. We adopt here a central cell potential which takes into account the incomplete screening of the Coulomb potential at very short distances from the dopant~\cite{SaraivaJPCM2015} 
\begin{equation}
V_{\rm cc}=\frac{2(\epsilon_s-1)e^{-r/r_{\rm cc}}}{r} \, ,
\label{eq:vcc}
\end{equation}
with $r_{\rm cc}$ a semiempirical parameter~\cite{SaraivaJPCM2015} calculated such that for a given acceptor the measured bulk ground state energy is reproduced, see Table~\ref{table:rcc}. A single $r_{\rm cc}$ characteristic of each dopant species is sufficient to get also the excited spectrum. The values are very similar for Si and Ge. However, due to the much smaller binding energies of the acceptor states in Ge (which corresponds to much more extended wave functions) the effect of the central cell correction on the binding energies is not as large in Ge as in Si. The central cell correction is not needed to reproduce the energy spectrum for the boron acceptor (namely, $r_{\rm cc}$ for boron is negligibly small). A larger binding energy corresponds to a larger $r_{\rm cc}$. Typical central-cell parameter values are very small, $r_{\rm cc} \sim 1$ \AA, and hence we do not expect it to be affected by the presence of the interface.

\begin{table}[h!]
\centering
\setlength{\tabcolsep}{6pt}
\begin{tabular}{c c c c c c c}
\hline \hline
  & & B &Al & Ga & In & Tl \\ \hline
  & E$_{\rm GS}$ (meV) & $45.83$ &  $69.03$ & $74.16$ & $157$ & $246$ \\
   Si & & & & & & \\
 &$r_{\rm cc}$ (nm) & - & $0.078$ & $0.082$ & $0.12$ & $0.15$ \\ \hline
   & E$_{\rm GS}$ (meV) & $10.82$ &  $11.15$ & $11.32$ & $11.99$ & $13.45$ \\
 Ge & & & & & & \\
 &$r_{\rm cc}$ (nm)& -& $0.077$ & $0.089$ & $0.12$ & $0.16$ \\ \hline \hline
\end{tabular}
\caption{Central cell parameter $r_{\rm cc}$~\cite{LipariSSC1980} that reproduces the measured bulk ground state energy E$_{\rm GS}$ for the different acceptor species~\cite{madelungbook}. A single $r_{\rm cc}$ suffices to reproduce the full spectrum~\cite{SaraivaJPCM2015}. Boron binding energies are well reproduced without the central cell correction. }
\label{table:rcc}
\end{table}

The effect of the proximity to the interface is considered by including in the Hamiltonian the image charges that arise due to the dielectric mismatch between the host crystal and the barrier~\cite{CalderonPRL2006, CalderonPRB2010}. A $(001)$ interface is considered at a distance $d$ from the acceptor, see Fig.~\ref{fig:problem}. The total Hamiltonian is then
\begin{equation}
H_{\rm acceptor}=H_{\rm KL}+V_{\rm cc}-\frac{2Q'}{\sqrt{\rho^2+(z+2d)^2}} +\frac{Q'}{2(z+d)} \, ,
\label{Hamiltonian}
\end{equation}
where the third and fourth terms are the acceptor and the hole images respectively with $Q'=(\epsilon_{b}-\epsilon_{s})/(\epsilon_{b}+\epsilon_{s})$. $\epsilon_b$ is the barrier static dielectric constant~\cite{CalderonPRL2006}. Note that for an insulating barrier $Q'<0$, namely the acceptor image is attractive for holes, and hence an enhancement of the binding energy is expected. Typical barriers considered have very large gaps compared to the binding energies involved. Therefore, we will assume the bound hole encounters a hard wall at the interface. We neglect the electric field potential in order to focus on the effect of the symmetry reduction by the interface.
 
The Hamiltonian in Eq.~(\ref{Hamiltonian}) has the symmetry of the bulk crystal. Close to an interface, the symmetry of the crystal is reduced. Effective changes in the spectrum, similar to the ones caused by uniaxial strain with the splitting of the light-hole and heavy-hole subbands~\cite{HasegawaPR1963, BirWiley1974,FischettiJAP2003, ThompsonIEEE2006}, are expected. These are discussed in the next section.

\section{Symmetries and Variational method}
\label{sec:symmetries}

In bulk, the acceptor problem has cubic symmetry with inversion so the transformation elements form the ${O}_{h}$ group. The three irreducible representations $\Gamma_6$ and $\Gamma_7$ with dimension two and $\Gamma_8$ with dimension four are all allowed in this group. As a result, the acceptor states are doubly or four-fold degenerate. Due to the inversion symmetry, parity is conserved and the states can be separated in well defined parity states ($\Gamma_6^+$, $\Gamma_6^-$, $\Gamma_7^+$, $\Gamma_7^-$, $\Gamma_8^+$ and $\Gamma_8^-$). However, central cell effects break the inversion symmetry, reducing the symmetry of the system to the tetrahedral double group $\overline{T}_d$.  The $\overline{T}_d$ group also has irreducible representations $\Gamma_6$, $\Gamma_7$ and $\Gamma_8$ but in this case parity is not a good quantum number. A consequence of the inversion symmetry breaking is the appearance of weak transitions between states with nominally identical parity, which would be forbidden if inversion symmetry were actually preserved~\cite{BaldereschiPRB1974}. However, as the central cell corrections are only important at small distances from the acceptor (see Table~\ref{table:rcc}), their effect is very local and hence the parity can be in general considered a good quantum number. In the Hamiltonian Eq.~(\ref{Hamiltonian}), we neglect this symmetry breaking as we consider a central cell correction with spherical symmetry.

The selection rules of the Hamiltonian Eq.~(\ref{Hamiltonian}) can be obtained after examining the differential operators and the couplings between different sets of the pseudo-angular momentum $|J,J_z \rangle$. Being the atomic orbital angular momentum $\bf{L}$, we can define a total angular momentum $\bf{F}$ as $\bf{F}=\bf{L}+\bf{J}$~\cite{BaldereschiPRB1974, LipariSSC1978}. The  selection rules for the quantum numbers $\bf{F}$ and $F_z=L_z+J_z$, which can be obtained applying the Wigner-Eckart theorem to the terms of cubic symmetry, are
\begin{equation}
\langle {\bf F}',F'_z|H_{\rm acceptor}|{\bf F},F_z \rangle \propto \delta_{F'_z,(F_z+(0,\pm 4))} \, .
\label{eq:rules}
\end{equation}

We can use these selection rules to relate the quantum numbers $\bf{F}$ and $F_z$ to the irreducible representations of the $O_h$ (or $\overline{T}_{d}$) group. Doubly degenerate eigenstates of the cubic symmetry with $F_z=\pm 1/2 +4n$ transform under the group as $\Gamma_6$ states, while two-fold degenerate states with $F_z=\pm 3/2 +4n$ transform like $\Gamma_7$ symmetry states. The four-fold degenerate eigenstates of the cubic symmetric terms in the Hamiltonian correspond to the $\Gamma_8$ representation and can have any half-integer $F_z$, always according to the selection rules.

The description of the states is modified in the presence of an interface. The inversion symmetry is broken by the interface, here assumed to be in the $(001)$ direction, so the parity is clearly not a well defined quantum number. Moreover, in the semi-infinite space, the spherical harmonics do not form an orthogonal basis and therefore ${\bf L}$ can not be a well defined quantum number. An immediate consequence is that the total angular momentum $\bf{F}=\bf{L}+\bf{J}$ is not well defined and states with different $\bf F$ are not orthogonal to each other. However, the $z$ projection of the atomic angular momentum is associated to the $\varphi$ coordinate so $L_z$ is not affected by the presence of the interface and, as $F_z$ is the sum of $L_z$ and $J_z$, the selection rule Eq.~(\ref{eq:rules}) holds. In terms of symmetry, the tetrahedral double group $\overline{T}_d$  is reduced to the tetragonal group $\overline{D}_{2d}$. The $\Gamma_8$ symmetry is not allowed in the $\overline{D}_{2d}$ group but the irreducible representations $\Gamma_6$ and $\Gamma_7$ remain. This implies that Êclose to an interface the four-fold degeneracy of the $\Gamma_8$ states is broken into two doubly degenerate states with symmetries $\Gamma_6$ and $\Gamma_7$ respectively. This effect of symmetry reduction by the interface is analogous to the effect of uniaxially strained silicon in the $(001)$ direction~\cite{SunJAP2007, BroeckxPRB1987, WangJPCM2009}.

 The barrier potentials usually considered are much larger than the typical binding energies and hence a hard-wall boundary condition $\Psi(z\leq -d)=0$ is assumed for the wave function. The interface boundary condition implies that the spherical symmetry usually assumed for the bound hole variational wave-function in bulk is not valid and it is more appropriate to work in cylindrical variables with the z-axis perpendicular to the interface.
With the information of the symmetries and the selection rules we can define a hydrogen-like variational basis set 
\begin{equation}
\psi(\rho,z,\varphi,\alpha_i)=(z+d)z^{l'}\rho^{|L_z|} r^{n'}e^{-\alpha_i r+i L_z \varphi}|J,J_z\rangle \label{variationalwf} \,,
\end{equation}
where $l'=L-|L_z|$ and $n'=n-L-1$ with $n>L$. A set of different $\alpha_i$ values is considered. $\rho$ is the cylindrical variable $\rho=\sqrt{r^2-z^2}$. The $(z+d)$ prefactor assures that the basis set satisfies the hard wall boundary condition. Given an $L_z$, the quantum number $J_z$ is chosen such that $F_z$ belongs to the convenient symmetry. This basis set is truncated at a certain $L_{max}$ with $n_{max}=L_{max}+1$. $L_{max}$ is chosen such that  the condition for the energy $(E_{L_{max}}-E_{L_{max}-1})=0.1$ meV is fulfilled in bulk, which in silicon corresponds to $L_{max}=11$. The number of different $\alpha_i$ considered is not as determinant for the convergence as the value of $L_{max}$. For example, for the ground state energy of an acceptor in bulk silicon, including $\alpha_1=1$ and $\alpha_2=2$ with $L_{max}=11$ gives $(E_{L_{max}}-E_{L_{max}-1})=0.1$ meV. However, the excited states require a larger set of different $\alpha_i$ due to their different extensions. Adding $\alpha_3=0.5$ and $\alpha_4=0.25$ gives results with $(E_{L_{max}}-E_{L_{max}-1})<0.2$ meV for the first 8 states. For Ge, the set is truncated at $L_{max}=11$ for the $J=3/2$ states while for the $J=1/2$ states $L_{max}=6$ since the split-off band is less relevant in this case and the convergence is faster.

Since the basis set defined in Eq.~(\ref{variationalwf}) is not orthonormal it is necessary to consider the overlap matrix $S_{i,j}$
\begin{eqnarray}
S_{i,j}&=&\langle\psi(\alpha_i)|\psi(\alpha_j)\rangle \nonumber \\
H_{i,j}&=&\langle\psi(\alpha_i)|H_{acceptor}|\psi(\alpha_j)\rangle  \,,
\label{eq:overlaph}
\end{eqnarray}
and the problem becomes a generalized eigenvalue problem
\begin{equation}
H_{i,j}|\Psi\rangle=E S_{i,j}|\Psi\rangle \,.
\end{equation}

Most of the integrals used to obtain the matrix elements of $H_{i,j}$ and $S_{i,j}$ can be solved formally, as detailed in Appendix~\ref{integrals}.

\begin{figure*}
\includegraphics[clip,width=0.325\textwidth]{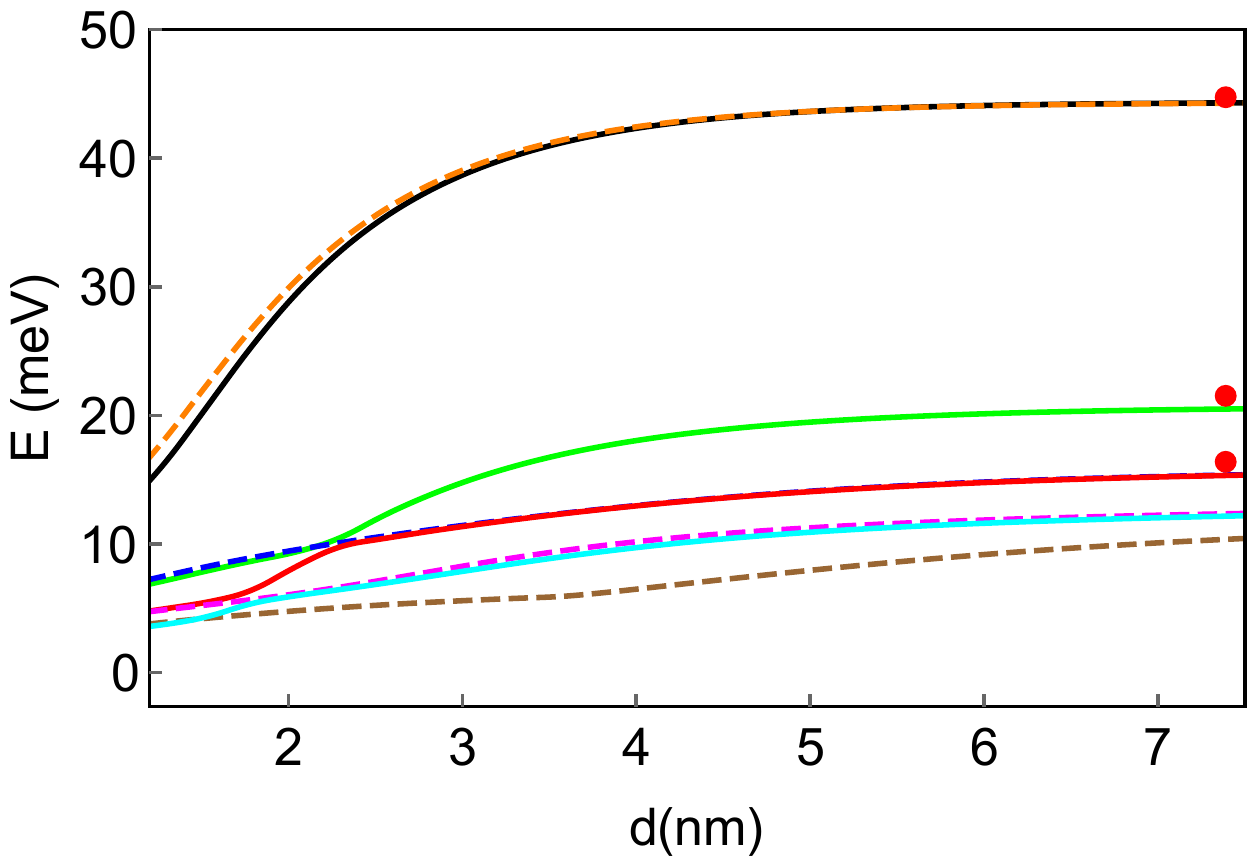}
\includegraphics[clip,width=0.325\textwidth]{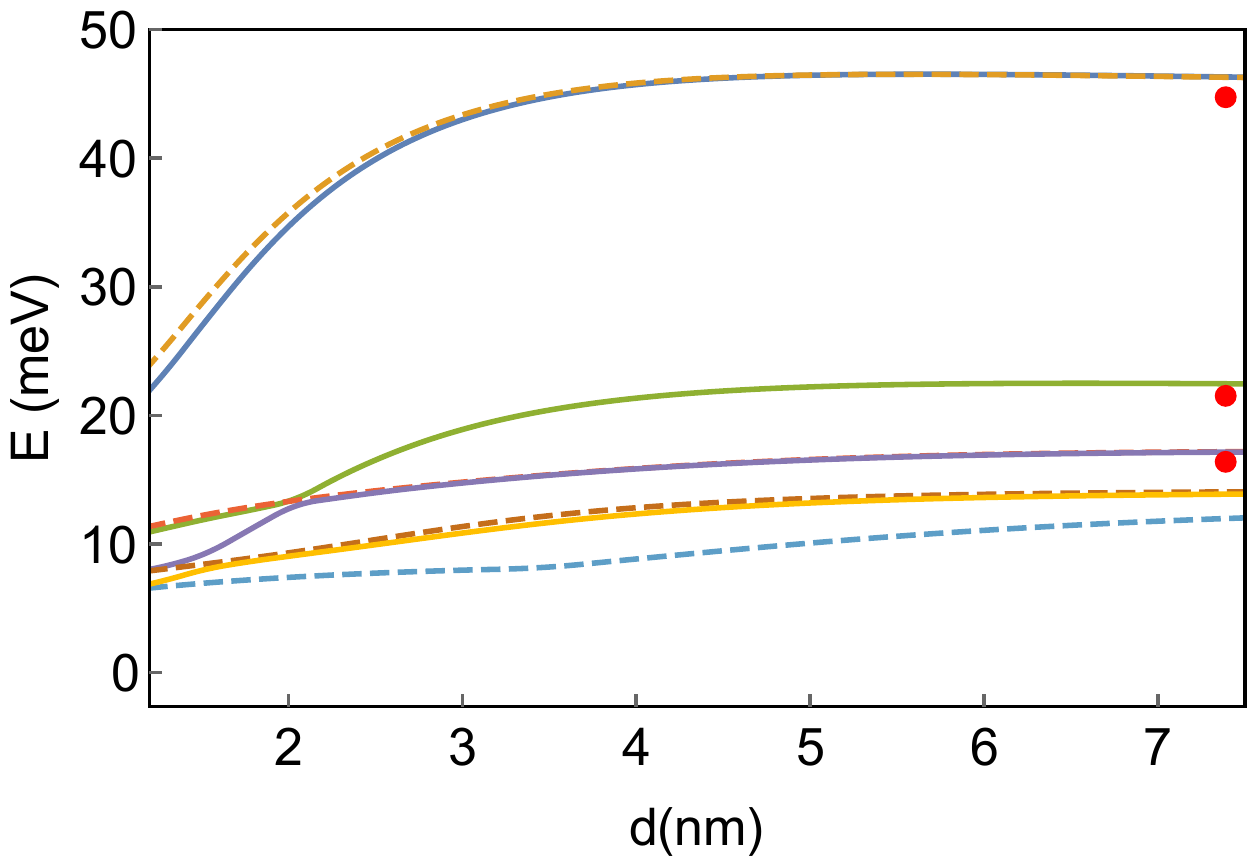}
\includegraphics[clip,width=0.325\textwidth]{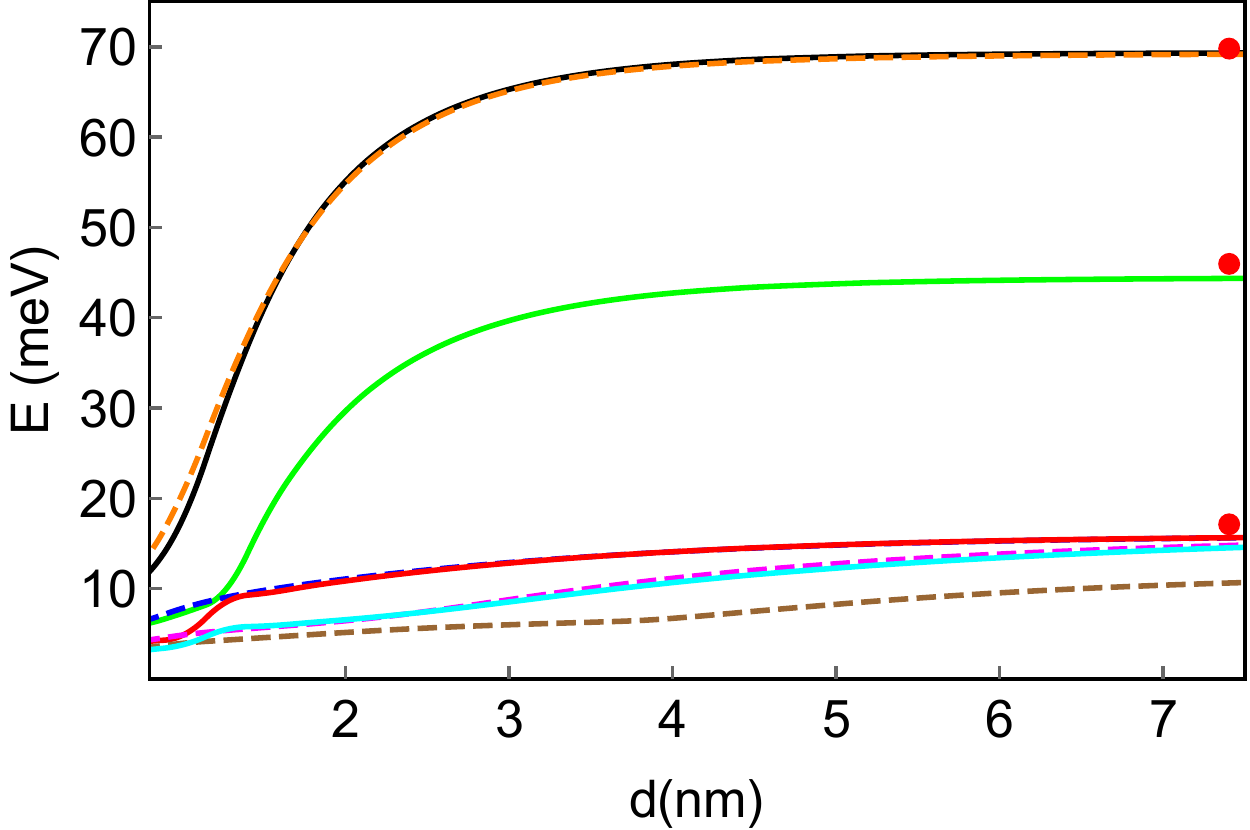}
\caption{\label{fig:spectrum} Energy dependence on distance to the interface of the ground state and some excited states of an acceptor in Si. The highest energy corresponds to the ground state which is four-fold degenerate in bulk and split in two Kramers doublets near the interface. Dashed lines indicate $\Gamma_7$ states while solid lines indicate $\Gamma_6$ states. The red dots are the experimental values for the bulk energies of the lowest three states~\cite{madelungbook,PajotSpringer2010}. As the excited states have a less localized wave function than the ground state, they become affected by the interface at larger distances. (a) For B acceptors, neglecting the image charges. (b) Same as (a) including the image charges corresponding to a SiO$_2$ barrier. Note the enhancement of the binding energies in (b) with respect to (a) due to the attractive character of the acceptor image. (c) Same as (a) for Al acceptors. Holes are more strongly bound for Al in bulk, see Table~\ref{table:rcc}, but the energies at small $d$ are very similar to the B case in (a). }
\end{figure*}

\begin{figure}
\includegraphics[clip,width=0.325\textwidth]{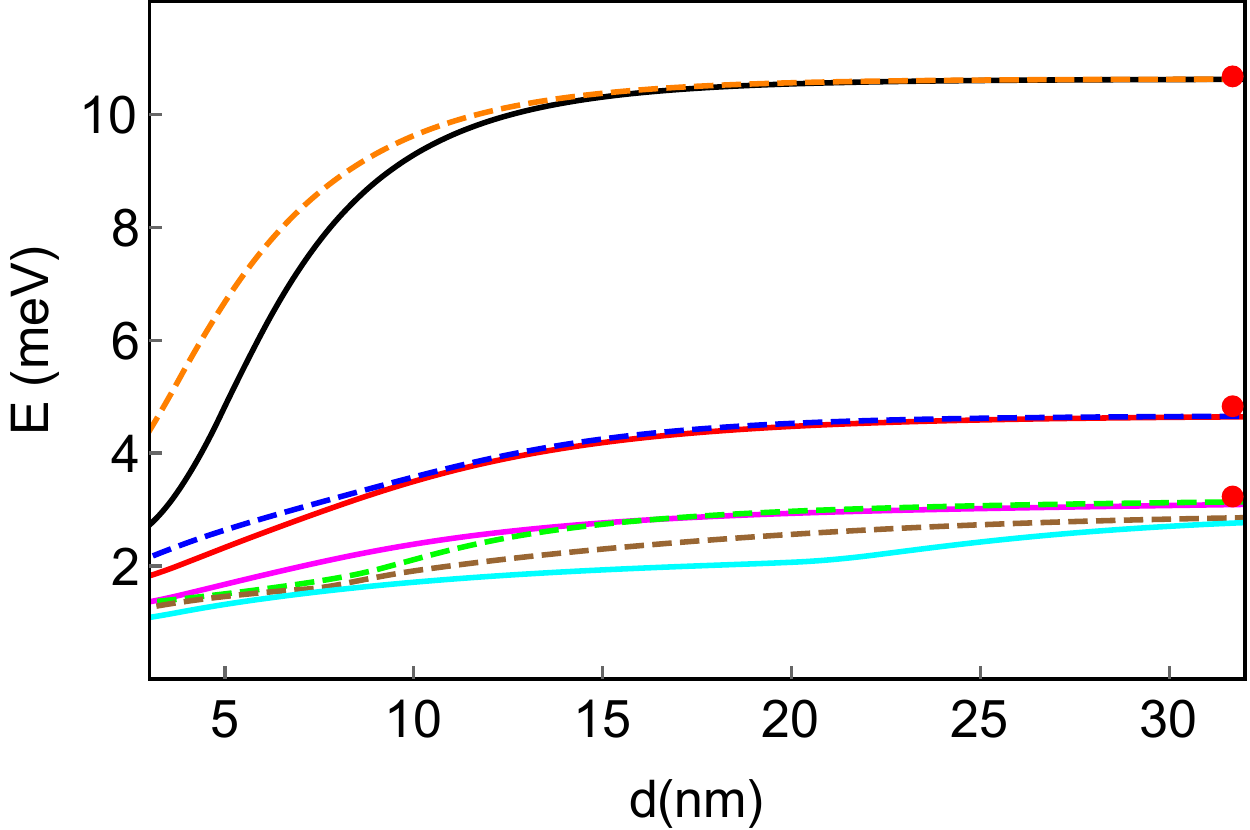}
\caption{\label{fig:Gespectrum} Spectrum of a B acceptor in Ge. The notation for the curves is the same as in Fig.~\ref{fig:spectrum}. The binding energy is much smaller in Ge than in Si, however, the ground state splittings are of the same order, see Fig.~\ref{fig:GevsSi-splitting}.}
\end{figure}

\section{Results and discussion}
\label{sec:results}

Fig.~\ref{fig:spectrum} illustrates the effect of the interface on the energy spectrum as a function of the interface-acceptor distance $d$ in Si. The superimposed dots at large $d$ correspond to the three lowest energies measured in bulk for the corresponding acceptors~\cite{madelungbook,PajotSpringer2010}. The bulk energies are well reproduced but the interface boundary condition and the image charges affect the calculated energies up to the distances shown in the figure. Fig.~\ref{fig:spectrum}~(a) corresponds to a B acceptor and neglects the image charges (namely, $Q'=0$), Fig.~\ref{fig:spectrum}~(b) considers a SiO$_2$ barrier ($Q'=- 1/2$), and Fig.~\ref{fig:spectrum}~(c) is the result for Al acceptors with $Q'=0$. Fig.~\ref{fig:Gespectrum} shows the corresponding results for B acceptors in Ge with $Q'=0$. There are two main qualitative interface induced effects on the energy levels: (i) the binding energies are smaller close to the interface and (ii) the ground state (which is four-fold degenerate in bulk) splits in two Kramers doublets~\cite{MolAPL2015}.

\begin{figure}
\includegraphics[clip,width=0.48\textwidth]{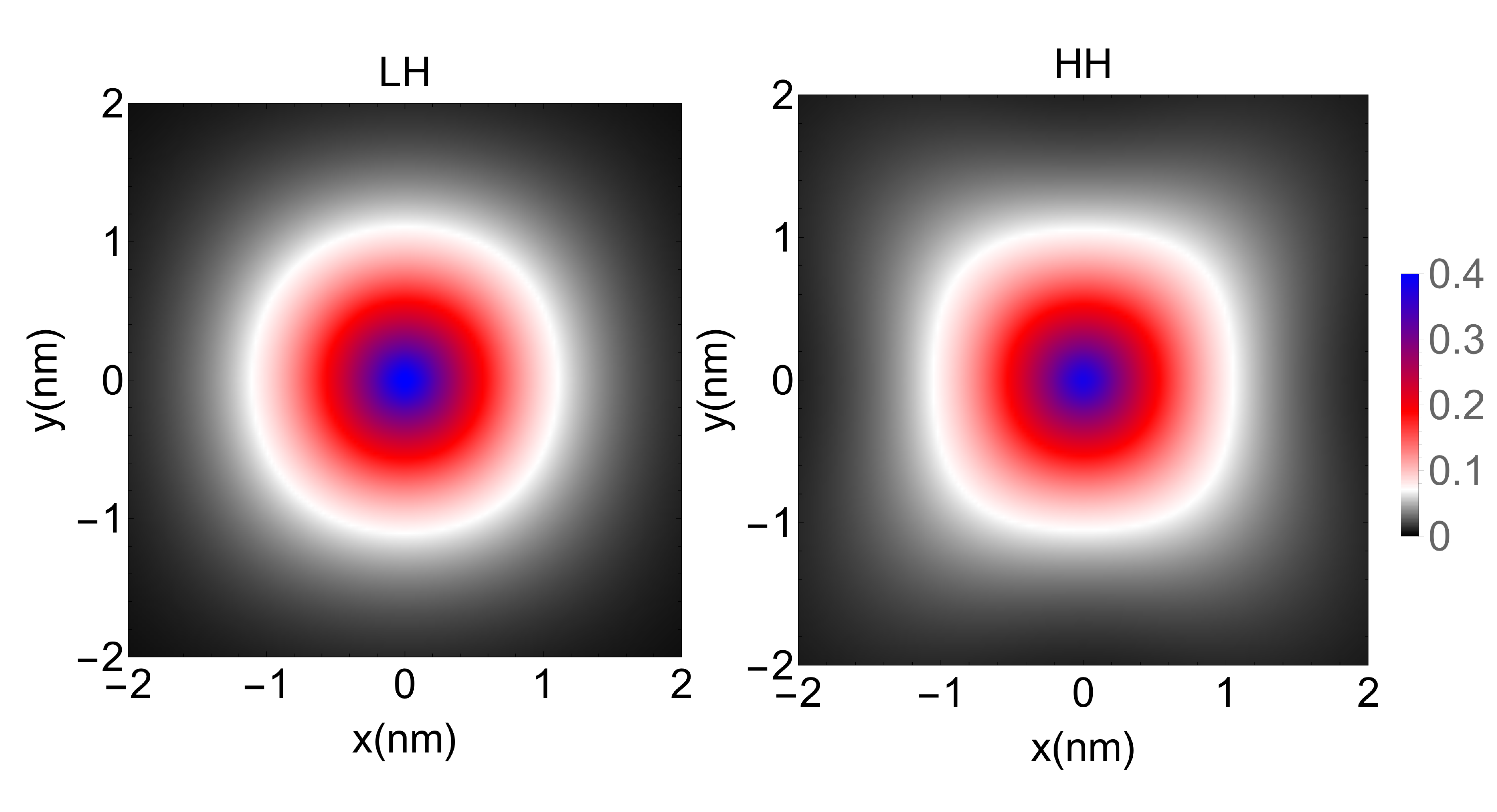}
\includegraphics[clip,width=0.48\textwidth]{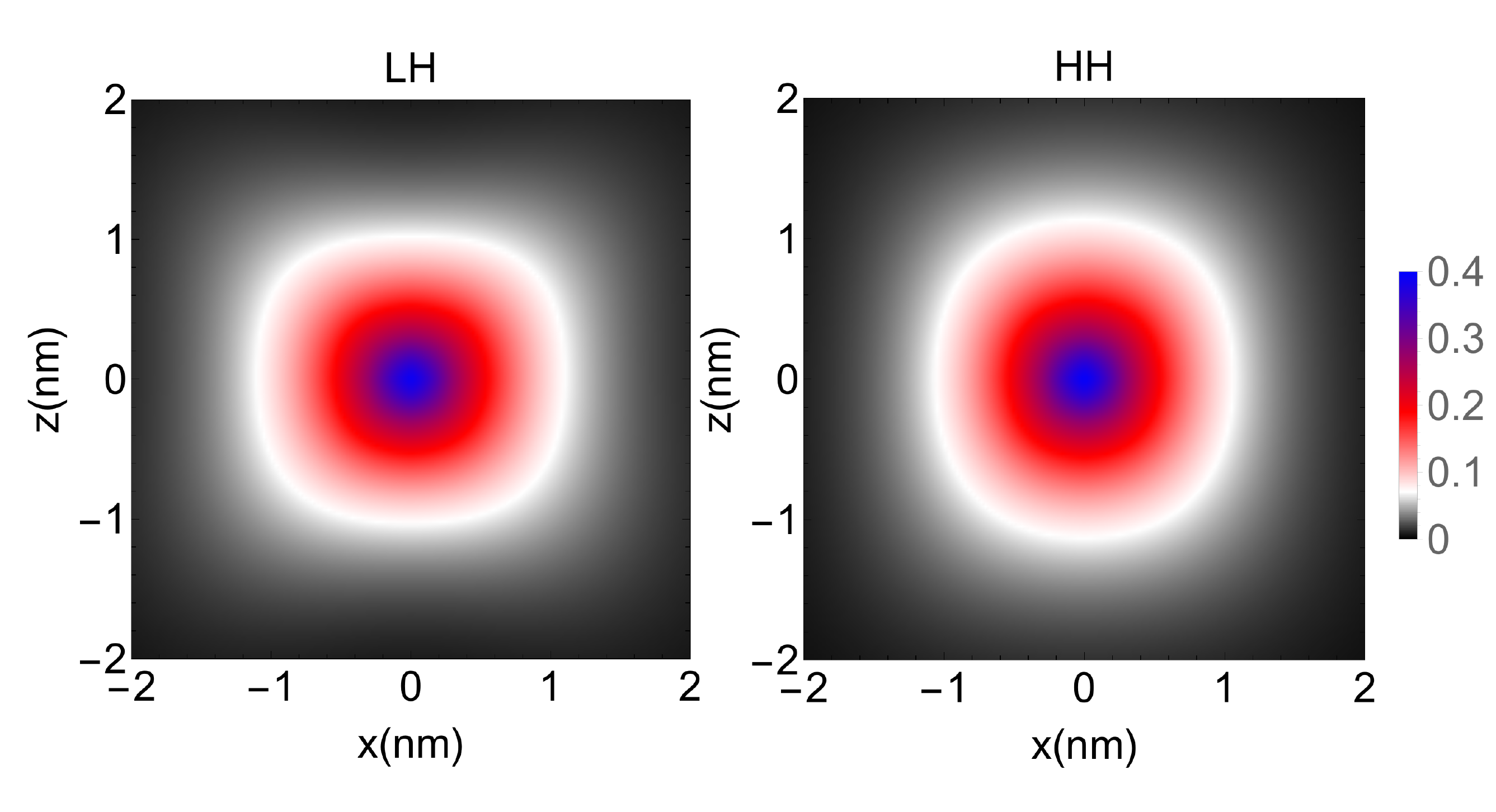}
\includegraphics[clip,width=0.48\textwidth]{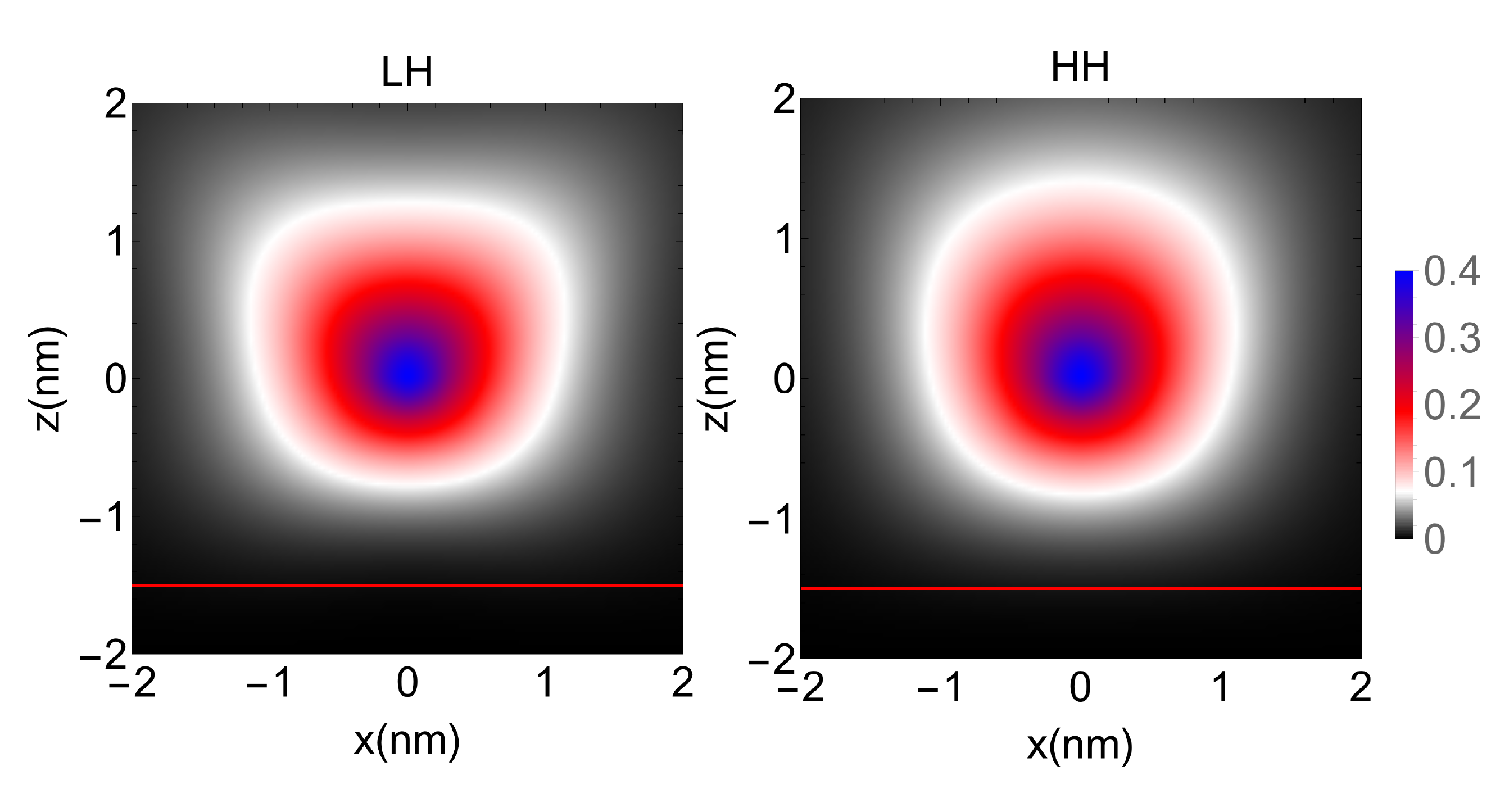}
\caption{\label{fig:wf} 
Spatial probability distribution of the two ground state doublets for B in Si considering a SiO$_2$ interface: (left) $\Gamma_6$ state with a predominant light-hole (LH) character, and (right) $\Gamma_7$ state with a predominant heavy-hole (HH) character. The acceptor is located at $(0,0,0)$. Top figures are the in-plane ($x-y$) images for $d=2$ nm. The shape of the wave-functions is the same at all distances. Differences with $d$ can be noticed in the $x-z$ plane (images are equivalent in the $y-z$ plane). The middle panels correspond to $d=7.5$ nm and the bottom ones to $d=1.5$ nm. The red line in the bottom panels represents the interface position. The wave-functions are deformed by the interface proximity. The LH wave-function is more affected than the HH one leading to the energy splitting. }
\end{figure}

The {\em reduction of the binding energies} close to the interface is due to the quantum confinement~\cite{delerue-lannoo,CalderonPRB2010,MolPRB2013} produced by the boundary condition on the wave function, which has to be zero at the interface. The wave-function is hence deformed with its probability density shifting away from the interface, see Fig.~\ref{fig:wf}. This effect is more significant for the levels coming from the four-fold degenerate bulk ground-state than for the excited states leading to the compression of the full energy spectrum~\cite{KovalenkoEA1991}. This compression also appears in bulk strained systems~\cite{BroeckxPRB1987,WangJPCM2009} due to the splitting of the heavy-hole and the light-hole bands.

The reduction of the ground state binding energy due to the quantum confinement is partially compensated by the dielectric mismatch with the insulating barrier~\cite{MolPRB2013}, compare panels (a) and (b) in Fig.~\ref{fig:spectrum}: the holes are more strongly bound when the image charges are included (because $Q'<0$). The extra binding effect of the acceptor image can be still appreciated at the longest distances shown in Fig.~\ref{fig:spectrum} (b) by comparing to the bulk values. A vacuum barrier, with $Q'=-0.84$, would further increase the ground state binding energy such that, for $d=2$ nm, $E_{\rm GS} = 40.3$ meV, consistent with the reported experimental values in Ref.~\cite{MolPRB2013}. 

For doped Si, the energy difference between the two doublets and the excited spectrum can be lowered (for $Q'=0$) to values $< 8$ meV, a significant reduction from the $>25$ meV typical splitting in bulk. This energy splitting is enhanced when the dielectric mismatch is included, as in Fig.~\ref{fig:spectrum}~(b), but still smaller than its bulk value. The compression of the spectrum should be kept in mind when interpreting experimental measurements of bound states in field effect transistor geometries where SiO$_2$ is a common barrier material.

Qualitatively similar results are found for Al acceptors in Fig.~\ref{fig:spectrum} (c). In the bulk limit, the energies of the two first levels are enhanced with respect to B by the central cell corrections. However, the third level has a binding energy very similar to that in B. This difference is due to the first 2 energy levels being s-like (and hence more affected by central cell effects). Notoriously, although the Al acceptors have much larger binding energies in bulk, close to the interface the values are very similar to B acceptors. This is a consequence of the hole probability density shifting away from the dopant, significantly reducing the effect of the central cell correction on the energy. Therefore, distinction among different acceptors in terms of measured binding energies may be blurred by the proximity to an interface.

The {\em splitting of the ground state} is due to the symmetry reduction produced by the interface. As explained in Sec.~\ref{sec:symmetries}, the $\Gamma_8$ symmetry is not allowed and hence the states with $\Gamma_8$ symmetry in bulk acquire a $\Gamma_6$ or $\Gamma_7$ symmetry as $d$ is reduced. This is the case in particular of the four-fold degenerate bulk ground state which is split in two doublets. The two doublets have a predominant light-hole ($\Gamma_6$) or heavy-hole ($\Gamma_7$) character. Both corresponding wavefunctions have s-like envelopes, however, the light-hole ground state is more affected by the interface because it has a higher contribution of high order spherical harmonics parallel to the surface as shown in Fig.~\ref{fig:wf}. The slightly different shapes of the heavy-hole and light-hole wave functions leads to the energy splitting of the two doublets~\cite{MolAPL2015}. 

The solid and dashed lines in Fig.~\ref{fig:spectrum} correspond to $\Gamma_6$ and $\Gamma_7$ symmetries respectively. Whenever those curves are degenerate towards the bulk (increasing $d$), the $\Gamma_8$ symmetry is recovered. The level crossings (anticrossings) in the excited spectrum occur between states with different (same) symmetry. For small values of $d$ there are some near degeneracies between $\Gamma_6$ and $\Gamma_7$ states which are accidental and not related to the (prohibited) $\Gamma_8$ symmetry.

\begin{figure}
\includegraphics[clip,width=0.325\textwidth]{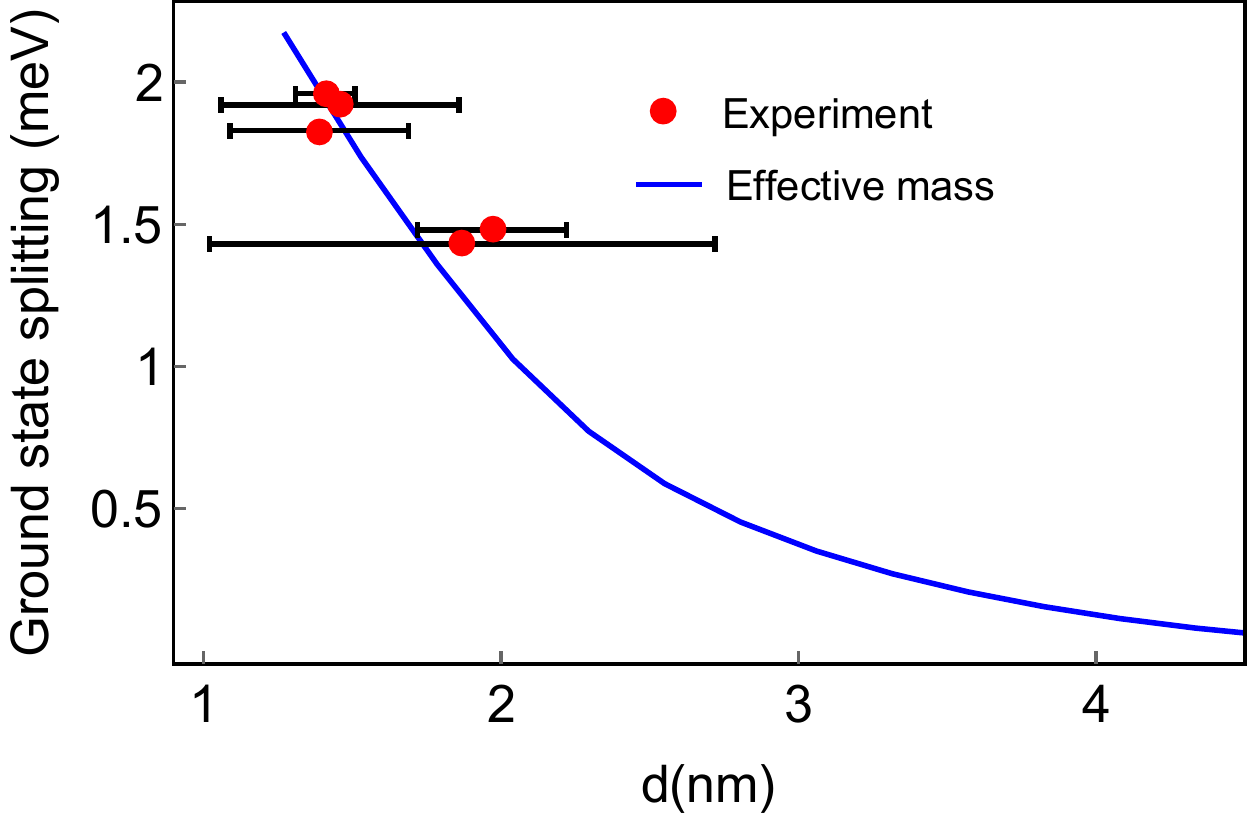}
\caption{\label{fig:Sisplitting} Splitting of the B ground state in Si as a function of the distance $d$. The red dots are the experimental values taken from Ref.~\cite{MolAPL2015}. Although an interface with vacuum has been considered in this plot ($Q'=-0.84$), the value of the splitting is basically independent of $Q'$ down to the distance $d$ considered here.}
\end{figure}

\begin{figure}
\includegraphics[clip,width=0.325\textwidth]{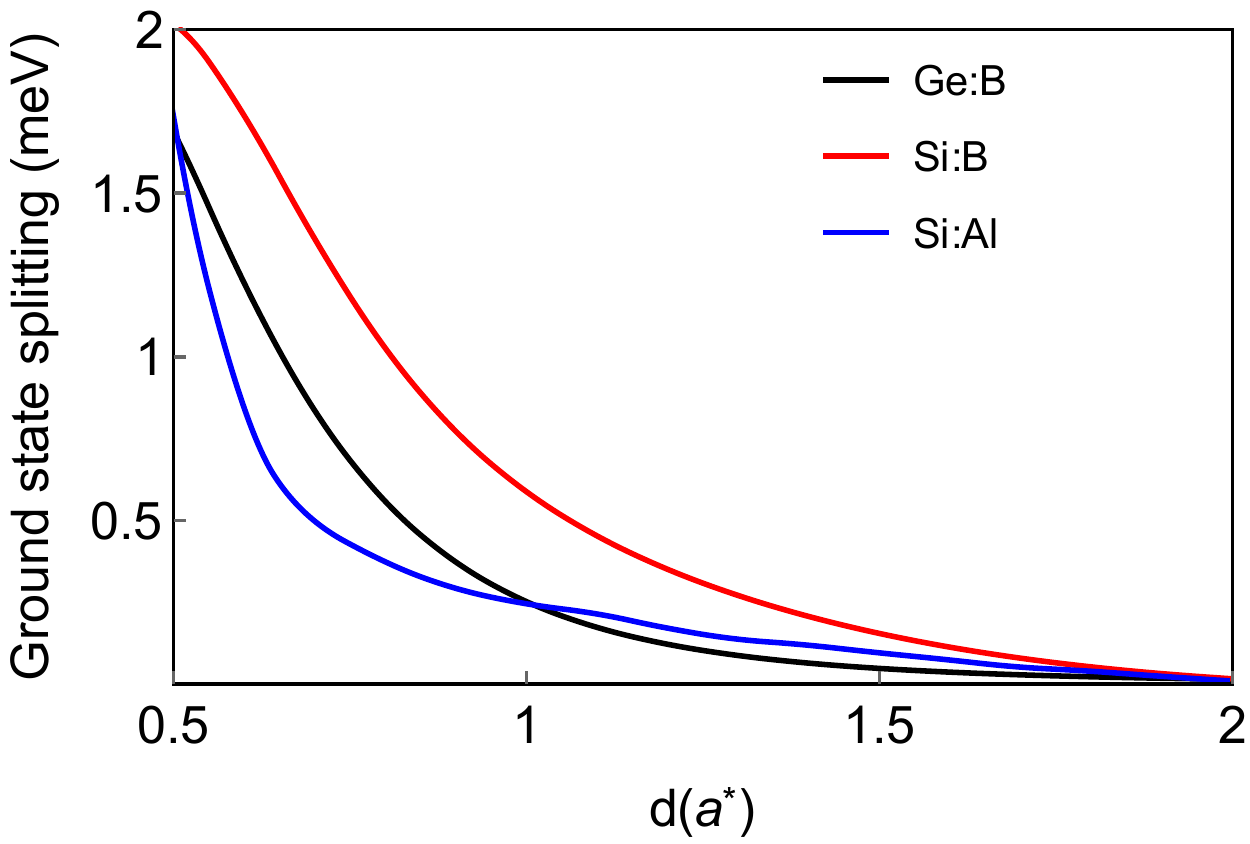}
\caption{\label{fig:GevsSi-splitting} Comparison of the splitting of the ground state in Si (with two different acceptors) and Ge in effective units of distance. }
\end{figure}

\begin{figure}
\includegraphics[clip,width=0.37\textwidth]{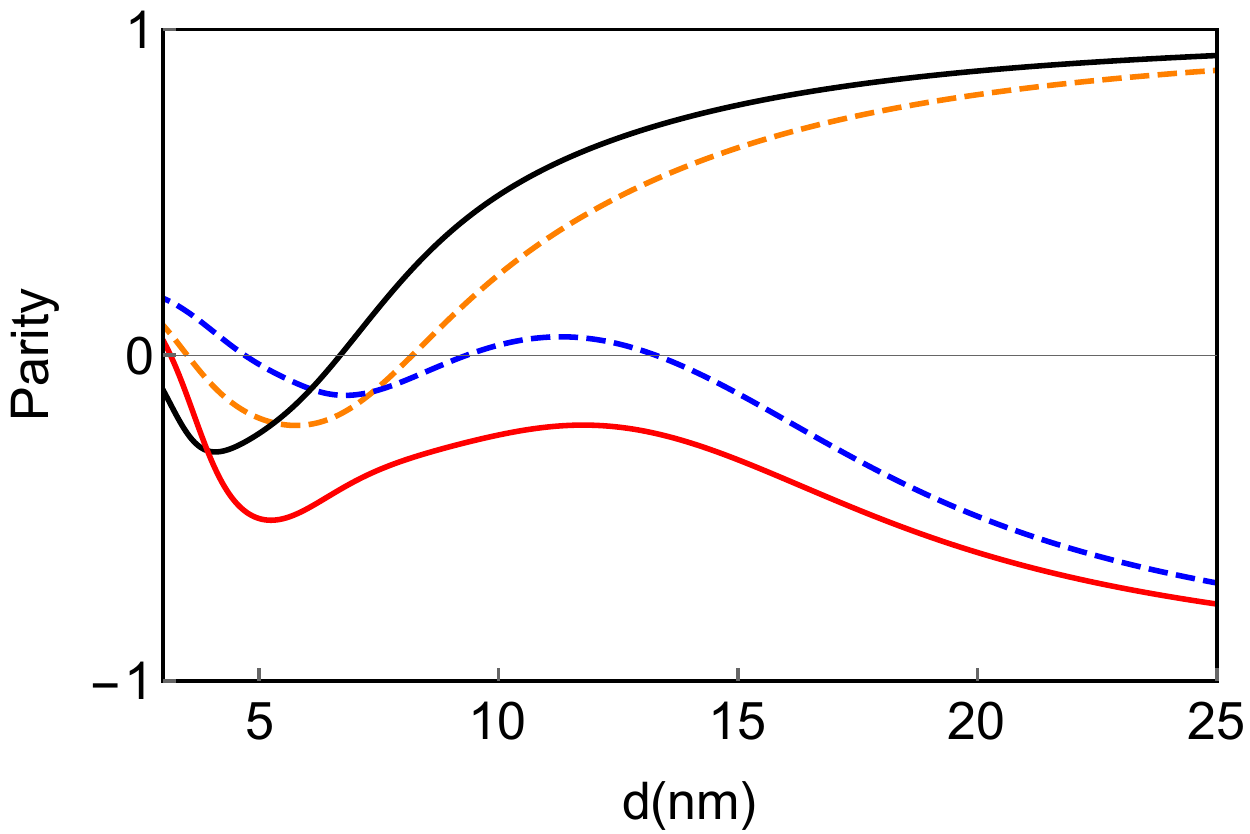}
\caption{\label{fig:parity} This figure illustrates how the parity is lost in the proximity of an interface for B in Si and neglecting the image charges ($Q'=0$). The color code for the lines is the same as in Fig.~\ref{fig:spectrum}: The solid lines indicate the $\Gamma_6$ states while dashed lines indicate $\Gamma_7$. The black and orange lines are the first two even states (bulk ground state). Red and blue are the first two odd states.}
\end{figure}

Fig.~\ref{fig:Sisplitting} shows the energy difference between the two lowest doublets for B in Si. The energies for this plot have been obtained considering an interface with vacuum ($Q'=-0.84$) but results are independent of the value of $Q'$ considered. The energy difference is enhanced as the acceptor gets closer to the interface. A decreasing splitting is found for distances $d_{\rm min} \sim 0.5 a^*$, smaller that the ones shown in the figures, probably signalling a qualitative change in symmetry similar to the one found in the limit of infinite strain, with extra degeneracies~\cite{BroeckxPRB1987,WangJPCM2009}. However,  we do not expect the effective mass approximation used here to be valid for values $d<d_{\rm min}$. 

The dots with error bars in Fig.~\ref{fig:Sisplitting} are experimental values as reported in Ref.~\cite{MolAPL2015}. The agreement with the calculated splitting is very good even in comparison to more sophisticated tight-binding calculations~\cite{MolAPL2015}. The giant splitting $\sim 6$ meV found at very short distances $d\sim 0.5$ nm $<d_{\rm min}$ in Ref.~\cite{MolAPL2015} is not included in the figure. We stress that the doublet splitting does not depend on $Q'$, namely it is independent of the nature of the insulator, but the energy spectrum compression does, see discussion of Fig.~\ref{fig:spectrum}. Therefore, in order to keep the excited states away from the lowest doublet in case of a giant splitting, an insulator with a very low dielectric constant should be used.

The same considerations can be made in the case of doped Ge, see Fig.~\ref{fig:Gespectrum} for the energy spectrum. The main difference with Si is that holes are much less bound in Ge, leading to a larger effective Bohr radius and hence the effect of the interface is notorious for much deeper acceptors. 

Although the energy spectrum is quantitatively affected by the host crystal and the acceptor species, the interface induced ground state splitting is very similar in different systems, as illustrated in Fig.~\ref{fig:GevsSi-splitting}. Here the doublet splitting is shown for B in Si, Al in Si and B in Ge as a function of the distance in effective units. The three curves are very similar and in all cases splittings $\gtrsim 1$ meV can be found. In comparison, the electric field required to achieve this splitting is much larger than the ionization field $5$ MV/m~\cite{SmitPRB2004} justifying the omission of the electric field in our model. 

As explained in Sec.~\ref{sec:symmetries}, breaking the inversion symmetry implies that the parity of a state is not a conserved quantity anymore. Fig.~\ref{fig:parity} quantifies (for the particular case of B in Si) how the parity of the first two even and two odd states lose their well defined parities. The wave-function $\Psi$ is a combination of the basis states $\psi$ in Eq.~\ref{variationalwf}, each with a well defined parity $P=\pm 1$. The parity in the figure is defined as $\langle \Psi|P|\Psi\rangle / \langle \Psi |\Psi\rangle$.  Even at the longest distances shown, $d=25$ nm, the different states are not completely even or odd. This implies that states with different parities have a finite overlap ($S \sim 5\%$ already at $d=25$ nm). $S$ increases strongly with decreasing $d$, reaching $S\gtrsim0.2$ at $d\sim2$ nm. Consequently, optical transitions which are not allowed in bulk, may be permitted  close to an interface.

\section{Summary and Conclusions}

We have used an effective mass approach to study the effect of an interface on the group III acceptor energy spectra and bound states symmetry in Si and Ge. This method, though computationally less demanding than tight-binding, has been proven to be very reliable not only for the calculation of binding energies but also for the wave-functions~\cite{SaraivaArXiv2015}. A semiconductor/insulator interface introduces a specific boundary condition and the corresponding dielectric mismatch implies new attractive potential terms in the Hamiltonian. The combined effect of the quantum confinement and the dielectric mismatch gives rise to energy shiftings which depend on the distance between the acceptor and the interface and the value of the insulator dielectric constant. In general terms the full energy spectrum is compressed, namely, the distance between the ground state and the excited states is reduced by the confinement but this reduction is partially compensated by the dielectric mismatch~\cite{MolPRB2013}.  Central cell effects, which account for the binding energies dependance on the acceptor species, become less important when acceptors get closer to the interface with the insulator. Therefore, different acceptors close to an interface may be difficult to distinguish by the values of their binding energies.

We have also followed the modifications on the symmetry of the bound states which can be qualitatively understood via the analysis of the symmetry breaking induced by the interface. One of the consequences of this symmetry reduction is the splitting of the four-fold ground state in two Kramers doublets, as reported in Ref.~\cite{MolAPL2015}. This doublet splitting is independent of the dielectric mismatch for a particular host:acceptor combination. Different acceptors lead to comparable values of the interface induced doublet splitting, which is typically $\gtrsim 1$ meV. Our results are in very good agreement with the measurements  in Ref.~\cite{MolAPL2015}. Inversion symmetry breaking implies that the parity is not a good quantum number leading to parity mixing and allowing for optical transitions which are not allwed in bulk. 

It will be useful to keep in mind these results in the implementation of an acceptor based quantum computer. The strong dependence of the binding energies, particularly for the ground state, on the distance between the acceptors and the interface introduces an uncertainty in the sample preparation due to the difficulty in controlling the positioning of dopants to the required degree. The interface induced ground state splitting could anyway be exploited after {\em a posteriori}Ê characterisation of the samples. Strain could provide a way to induce the ground state splitting if acceptors are buried farther from the interface to avoid acceptor to acceptor binding energy variability. Strain would also be more advantageous than interface effects if optical transitions are involved, as the former does not produce the parity mixing we find for subsurface acceptors. In any case, the fact that the ground state splits due to interface proximity allows the possibility of a tunable two-level system in state-of-the-art devices.

\acknowledgements
The authors thank fruitful discussions with Belita Koiller and Andr\'e Saraiva and acknowledge support from MINECO-Spain through Grant FIS2012-33521. JCAU thanks the support from "Ayudas para contratos predoctorales para la formaci\'on de doctores 2013", grant BES-2013-065888.

\appendix
\renewcommand{\thefigure}{A\arabic{figure}}
\renewcommand{\thetable}{A\arabic{table}}
\setcounter{figure}{0}  
  
\begin{widetext}
\section{Integrals \label{integrals}}
The calculation of the matrix elements $H_{i,j}$ and $S_{i,j}$ have required the evaluation of integrals involving products of wavefunctions (\ref{variationalwf}) and the expected values of the different operators (\ref{PQLM}). The analytical solutions to these integrals are not tabulated. We summarize here most of the integrals used in this work.

The general form of the integrals involved in the calculations of the matrix elements is

\begin{equation}
I(c,n,k')=\int_{-d}^{\infty}dz \int_{|z|}^{\infty}dr\ z^c r^n \left(\sqrt{r^2-z^2}\right)^{\ k'}e^{-\alpha r} \,.
\end{equation}

The evaluation of this integral depends strongly on the parity of the exponent $k'$. When $k'=2k$ being $k$ any positive integer and defining for convenience $\gamma=2k+n+c+2$:
\begin{equation}
I(c,n,2k)=\frac{1}{\alpha^{\gamma}}\left[\dfrac{k!\Gamma(\frac{c+1}{2})}{2\Gamma(\frac{c+3}{2}+k)}\left(\Gamma(\gamma)(1+(-1)^c)\right)+(-1)^{c+1}\Gamma(\gamma,\alpha d)\right]  \, ,
\end{equation}
being $\Gamma(a,z)$ the incomplete gamma function.

 When $k'$ is an odd number $k'=2k+1$, the parity of the exponent $n$ becomes relevant. If $n$ is even
\begin{eqnarray}
I(c,n,k')=\sum_{m=0}^k(-1)^m\sum_{u=1}^{\frac{\gamma-c}{2}-m}(\prod_{l=0}^{u-1}2l+1)\binom{k}{m}
\frac{1}{\alpha^u}\binom{n/2+k-m}{u-1}\Big[\frac{2^{\gamma-u-1}}{\alpha^{\gamma+u+1}}\Gamma(\frac{\gamma+1}{2}) \nonumber \Gamma(\frac{\gamma+1-2u}{2}) \\ +(-1)^c2^{-u-2}d^{\gamma+1-u}  (\alpha d)^{-u}\pi \csc(u\pi)\Big(4^u\Gamma(\frac{\gamma+1-2u}{2}) \nonumber _1\tilde{F}_2(\frac{\gamma+1-2u}{2};1-u,\frac{\gamma+3-2u}{2};\frac{\alpha^2 d^2}{4}) \nonumber \\ -(\alpha d)^{2u}\Gamma(\frac{\gamma+1}{2}) _1\tilde{F}_2(\frac{\gamma+1}{2};\frac{\gamma+3}{2},u+1;\frac{\alpha^2 d^2}{4})\Big)\Big] \, ,
\end{eqnarray}
being $\tilde{F}$ the hypergeometric regularized function. And when n is odd
\begin{eqnarray}
I(c,n,k')=\sum_{m=0}^k(-1)^m\sum_{u=1}^{\frac{\gamma+1-c}{2}-m}(\prod_{l=0}^{u-1}2l+1)\binom{k}{m}
\frac{1}{\alpha^u}\binom{n/2+k-m-1/2}{u-1}\Big[\frac{2^{\gamma-u-1}}{\alpha^{\gamma+u+1}}\Gamma(\frac{\gamma+2}{2}) \nonumber \Gamma(\frac{\gamma-2u}{2}) \\ +(-1)^c2^{-u-3}d^{\gamma+1-u}  (\alpha d)^{-u-1}\pi \csc(u\pi) \Big(4^u\Gamma(\frac{\gamma-2u}{2}) \nonumber _1\tilde{F}_2(\frac{\gamma-2u}{2};-u,\frac{\gamma+2-2u}{2};\frac{\alpha^2 d^2}{4}) \nonumber \\ -(\alpha d)^{2u+2}\Gamma(\frac{\gamma+2}{2}) _1\tilde{F}_2(\frac{\gamma+2}{2};\frac{\gamma+4}{2},u+2;\frac{\alpha^2 d^2}{4})\Big)\Big]  \, .
\end{eqnarray}

The exponent $n$ can be negative for certain operators. When this is the case, the integrals can be transformed into the previous integrals by using the method of differentiation on  $\alpha$ under the integral sign. For example, when $n=-1$
\begin{equation}
I(c,-1,k')=-\int d\alpha\ I(c,0,k')+C  \, ,
\end{equation}
where $C$ is a constant that can be obtained using $I(c,n,k)=0$ when $\alpha\rightarrow\infty$. The rest of negative $n$ integrals can be obtained using this method recursively.

\end{widetext}

\bibliography{acceptors}

\end{document}